\documentclass{jltp} 
\usepackage{graphicx}
\usepackage{amsmath}
\usepackage{amssymb}
\usepackage{bm}

\bmdefine\bv{v}
\bmdefine\br{r}
\bmdefine\bomega{\omega}
\bmdefine\bOmega{\Omega}
\bmdefine\bnabla{\nabla}
\bmdefine\bkappa{\kappa}

\begin{document}

\def\topfraction{1} \def\textfraction{0}

\title{\boldmath Phase diagram of turbulence in superfluid $^3$He-B}

\author{A.P.~Finne$^{*}$, S.~Boldarev$^{*,\dagger}$, V.B.~Eltsov$^{*,\dagger}$, and
M.~Krusius$^*$}

\address{$^*$Low Temperature Laboratory, Helsinki University of
  Technology,\\ P.O.
  Box 2200, FIN-02015 HUT, Finland\\
  $^\dagger$Kapitza Institute for Physical Problems, 119334 Moscow, Russia}
\vspace{-6mm}

\runninghead{A.P.~Finne \textit{et al.}}{Turbulence in superfluid $^3$He-B}

\maketitle

\vspace{-6mm}

\begin{abstract}
  In superfluid $^{\it 3}\!$He-B mutual-friction damping of vortex-line
  motion decreases roughly exponentially with temperature. We record as a
  function of temperature and pressure the transition from regular vortex
  motion at high temperatures to turbulence at low temperatures.  The
  measurements are performed with non-invasive NMR techniques, by injecting
  vortex loops into a long column in vortex-free rotation. The results
  display the phase diagram of turbulence at high flow velocities where the
  transition from regular to turbulent dynamics is velocity independent. At
  the three measured pressures 10.2, 29.0, and 34\,bar, the transition is
  centered at 0.52\,--\,0.59\,T$_c$ and has a narrow width of 0.06\,T$_c$
  while at zero pressure turbulence is not observed above~0.45\,T$_c$.

\end{abstract}

\section{INTRODUCTION}

According to the original Landau picture of superfluidity superflow is
irrotational: $\bnabla \times \bv_{\rm s} = 0$. This condition severely
restricts the possible motions of the superfluid: When placed in a rotating
container, it cannot participate in rotation. Later Onsager and Feynman
suggested the modern view of superfluid dynamics: By the creation of
quantized vortex lines the superfluid is able to mimic arbitrary complex
flows on length scales which are large compared to the inter-vortex
spacing. It was discovered more than 40 years ago that if superfluid $^4$He
is driven sufficiently fast then a complex vortex tangle appears. This
tangle produces apparently chaotic time-dependent flow at different length
scales.  Such motion was called superfluid or quantum
turbulence.\cite{VinenReview}

Studies of turbulence in a different superfluid, the B phase of
superfluid $^3$He, have started only
recently.\cite{Turbulence,AndreevReflection} From the hydrodynamics point
of view $^3$He-B differs from $^4$He in two important respects.
First, the viscosity of the normal component in $^3$He-B is large compared to
both the (sample size)$\times$velocity in a typical experiment and to the
circulation quantum. This means that the normal component of $^3$He-B is
always in well-defined externally imposed laminar motion, which
considerably simplifies the analysis of the experimental results. Second,
the damping of the superfluid motion varies significantly in $^3$He-B as a
function of temperature. Such damping comes from the interaction of the
thermal quasiparticles with the cores of the quantized vortices (so-called
mutual friction). The magnitude of the friction\cite{Bevan} changes from
large values (typical for superconductors) in the limit $T\rightarrow
T_{\rm c}$ to small values (typical for $^4$He-II) at temperatures $T <
0.5 \,T_{\rm c}$.

In a recent experiment\cite{Turbulence} we discovered that there is a sharp
transition in the character of superfluid motion in $^3$He-B as a
function of temperature at $P = 29\,$ bar pressure. When the laminar
vortex-free superflow is prepared by rotation of the cylindrical sample and
a few seed vortex loops are injected into it, then these loops expand in a
regular manner, with their number conserved at $T \gtrsim 0.6\, T_{\rm c}$
or, proliferate to a turbulent vortex tangle at $T \lesssim 0.6\, T_{\rm c}$.
(Eventually the turbulent tangle polarizes to mimic the global rotating
flow of the normal component which compensates the driving force and the
turbulence decays to a cluster of rectilinear lines containing $\sim 10^3$
vortices.)  

The transition to turbulence is controlled by the mutual
friction damping.  It can be understood by analyzing the
dynamic equation for the coarse-grained superfluid velocity $\bv_{\rm s}$
which is averaged over volumes containing many vortex lines:\cite{Sonin}
\begin{equation}
\frac{\partial \bv_{\rm s}}{\partial t}
+ \bnabla \biggl(\mu + \frac{v_{\rm s}^2}{2} \biggr) =
\bv_{\rm s} \times \bomega 
+ \alpha' \bomega \times (\bv_{\rm s} - \bv_{\rm n})
+ \alpha\, \hat\bomega \times [ \bomega \times (\bv_{\rm s} - \bv_{\rm n}) ].
\label{eq:dyn}
\end{equation}
Here $\bv_{\rm n}$ is the flow imposed on the normal component, $\bomega =
\bnabla \times \bv_{\rm s}$, $\hat\bomega$ is a unit vector in the
direction of $\bomega$, $\mu$ is the chemical potential, $\alpha$ and
$\alpha'$ are the dissipative and reactive mutual friction coefficients,
respectively. The vortex line tension is neglected. Applying simple
dimensional comparison of the magnitudes of the dissipative term (containing
$\alpha$) to the inertial terms in Eq.~\eqref{eq:dyn}, similar to how one
introduces the
Reynolds number in the classical Navier-Stockes equation,
one arrives to the conclusion\cite{Turbulence,LammiProc} that the character
of the flow is controlled by the ratio of the mutual friction coefficients
$q = \alpha / (1-\alpha')$: Turbulence occurs only if $q < q_{\rm c}\sim
1$. The transition from laminar to turbulent dynamics becomes
velocity-independent in the range of validity of Eq.~\eqref{eq:dyn} (which
includes the requirement for a sufficiently large flow velocity). This is seen
from the scale invariance of Eq.~\eqref{eq:dyn}: If $\bv_{\rm
  s}(\br,t)$ is the solution of this equation (and of the continuity
equation ${\rm div}\,\bv_{\rm s}=0$) for a given imposed flow $\bv_{\rm n}$,
then $\lambda\bv_{\rm s}(\br,\lambda t)$ is the solution for the scaled
imposed flow $\lambda\bv_{\rm n}$.

In this report we extend the investigation of the transition between
regular and turbulent vortex dynamics in $^3$He-B to the 
pressure range from zero up to the solidification pressure and examine the
question whether the transition to turbulence is controlled by a single
universal value $q_{\rm c}$ at all conditions.

\vspace{-5mm}
\section{EXPERIMENT}
A detailed account of the experimental techniques, including the injection
of the seed vortex loops and the NMR detection, is given in
Ref.~\onlinecite{LammiProc}. Here we present a brief summary. The $^3$He
sample is contained in a cylindrical container with radius $R=3\,$mm and
height 11 cm. At mid-height there is a barrier magnet which provides a
magnetic field to stabilize the A phase at all temperatures. The lower and
upper sections of the sample remain in the B phase. Two NMR detection coils
are installed in these sections close to the top and bottom ends of the
sample. From the NMR measurements the number of vortex lines within the
coil can be determined.

In the absence of the A phase the B phase in our experiments remains
vortex-free in rotation up to some container-specific critical angular
velocity.  Its value is not in good control but for the containers used in
the present work it exceeds 2\,rad/s at $T < 0.8\, T_{\rm c}$. When the A
phase is present in the middle of the sample the seed vortex loops are
injected into the B phase at a well defined and reproducible $\Omega$ as a
result of the Kelvin-Helmholtz (KH) instability of the AB interface.\cite{KH}
This velocity is a well-understood smooth function of $T$, $P$ and of the
current in the barrier magnet $I_{\rm b}$.  The circulation carried by the
initial B-phase vortices comes from the vortex layer which covers the AB
interface. In the layer vorticity is arranged in a regular manner, with
vortices aligned radially and parallel to the AB
interface.\cite{AB-VortexLayer} Thus one may expect that the configuration
of the seed vortices is reasonably well reproducible with one end of each
vortex sticking out of the AB interface and the other end perpendicular to
the outer wall of the sample. Our measurements indicate that the size of a
seed vortex is about 0.4\,mm.\cite{AB-VortexLayer} The injection parameter,
which is not in good control, is the number of the seed vortices, produced
by one instability event. This number has a rather wide distribution in the
range 3\,--\,30 vortices,\cite{LammiProc} with the average value of about
10. The reason for such a wide variation is presently not known.

The result from the expansion of the seed vortices along the B-phase column is
observed when they reach the NMR coils. A few vortices (in the case of
regular expansion) and an array with many vortex lines (in the case of
turbulent expansion) are easily distinguished at $T > 0.4\, T_{\rm c}$. The
two B-phase sections of the sample present two independent experimental
volumes. Their simultaneous measurement improves the statistics of
the results.

\vspace{-4mm}
\section{RESULTS}
\begin{figure}[t]
\centerline{\includegraphics[width=1.0\linewidth]{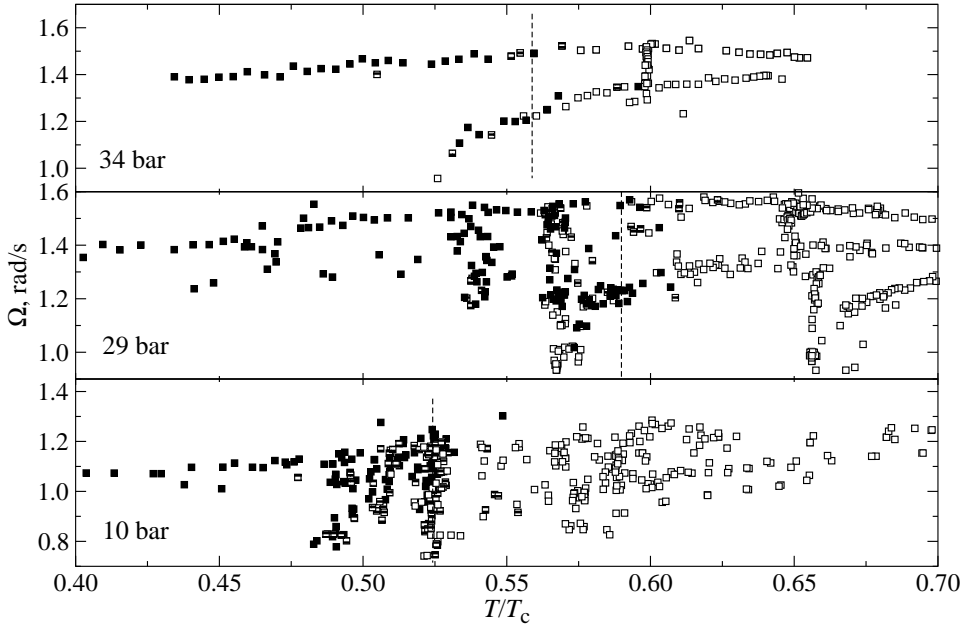}}
\caption{ Transition between regular (open squares) and turbulent 
  (filled squares) vortex dynamics as a function of temperature at three
  pressures. The dashed vertical lines represent the average values $T_{\rm
    t}$ of the
 distributions in Fig.~\protect\ref{TurbDistribution}.
 } \label{TurbTransition-vs-T}
\end{figure}

\begin{figure}[t]
\centerline{\includegraphics[width=1.0\linewidth]{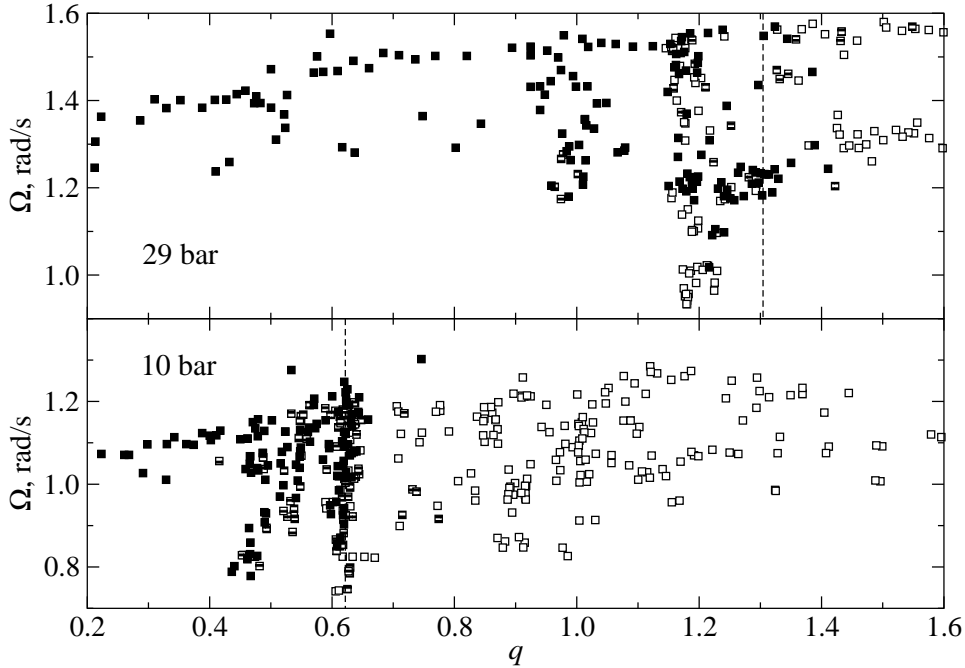}}
\caption{ Transition between regular (open squares) and turbulent 
  (filled squares) vortex dynamics as a function of the ratio of the
  mutual friction coefficients $q = \alpha / (1-\alpha^{\prime})$ at the two
  pressures where $q(T,P)$ was measured in Ref.~\onlinecite{Bevan}. }
\label{TurbTransition-vs-q}
\end{figure}

Figs.~\ref{TurbTransition-vs-T}\,--\,\ref{TurbTransition-vs-q} summarize
the transition from regular to turbulent dynamics with decreasing
temperature and trace the transition as a function of $\Omega $, $T$, and
$P$. Here $\Omega$ is the critical velocity for the KH instability, at
which vortices have been injected in the flow. In most cases the B phase
has been completely vortex-free before the injection and thus the flow in the
sample is $\bv_{\rm n} - \bv_{\rm s} = \bOmega \times \br$.  The data have
been measured along continuous trajectories, either by scanning temperature
at constant barrier current $I_{\rm b}$ or by scanning $I_{\rm b}$ at
constant $T$. These trajectories are not emphasized in the plots, instead
the data have been classified in regular (open symbols) or turbulent
(filled symbols) events, to highlight the transition separating them.

\begin{figure}[t]
\centerline{\includegraphics[width=1.0\linewidth]{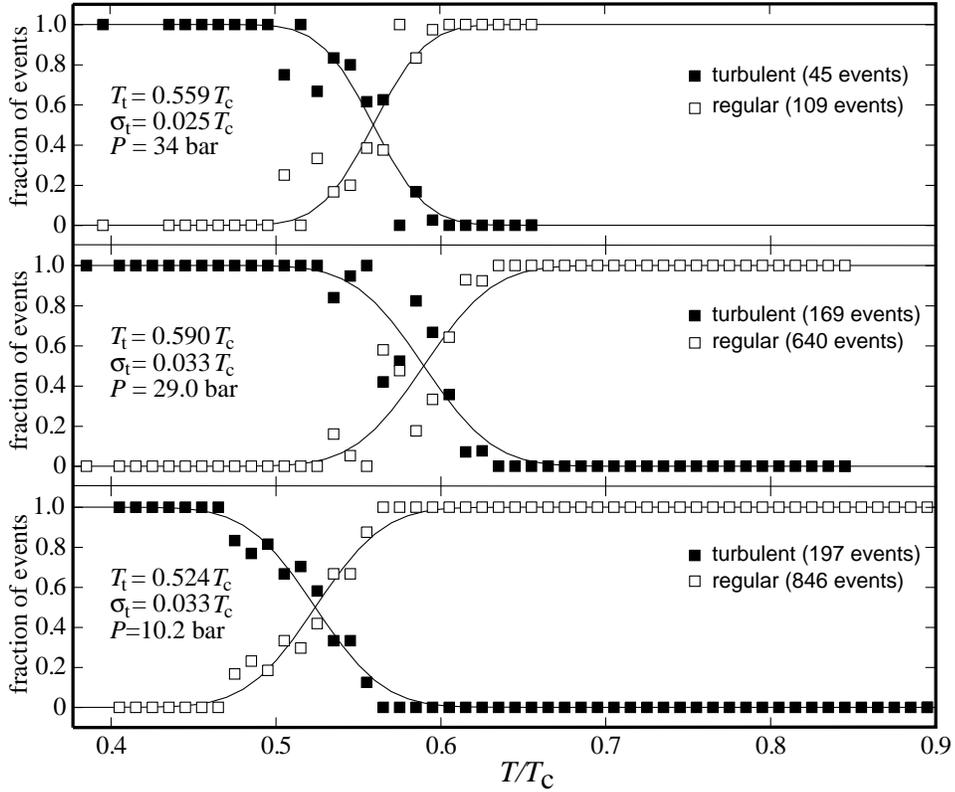}}
\caption{Probability for a vortex expansion event to be regular (open
  squares) or turbulent (filled squares) as a function of temperature at
  three pressures. The probabilities are calculated by arranging events
  from Fig.~\ref{TurbTransition-vs-T} as a function of temperature in bins
  of $0.01\,T_{\rm c}$ width at the three pressures. The continuous curves are
  fitted normal cumulative distribution functions with the average
  transition temperature $T_{\rm t}$ and dispersion $\sigma_{\rm t}$, as
  given in each panel.}
\label{TurbDistribution}
\end{figure}

On the basis of the data in Fig.~\ref{TurbTransition-vs-T} (which is
plotted in a different way in the later figures) we conclude that the
transition is independent of the velocity $\Omega$, if compared with the
width of the transition. This agrees with expectations based on the scale
invariance of the equation \eqref{eq:dyn}. This conclusion is valid for high
velocities, while low velocities much below the present range of $\sim
1\,$rad/s should be checked separately. 

Other features demonstrated by Fig.~\ref{TurbTransition-vs-T} are that the
transition has a certain width and that the transition temperature might
depend on pressure. To quantify these features the probabilities to oberve
regular and turbulent events are plotted in Fig.~\ref{TurbDistribution} as
a function of temperature for the 3 pressures.  Each of these distributions is
fitted with a normal cumulative distribution function with average value
$T_{\rm t}$ and dispersion $\sigma_{\rm t}$ as fitting parameters. This
function has been chosen as a simple way to obtain numerical values for
the average transition temperature and its width. We do not claim that the
usage of the Gaussian distribution has a solid physical reason.

For all 3 pressures the transition half-width is found to be around
$\sigma_{\rm t} \approx 0.03\,T_{\rm c}$. We believe that fluctuations in
the injection process are the main cause for the widths in
Fig.~\ref{TurbDistribution}. It appears probable that at higher
temperatures only some of the initial configurations of vortices (say, the
ones with a larger number of initial loops) will evolve to a turbulent
network. As the temperature decreases more of the different initial
configurations become unstable towards turbulence. This picture is
supported by comparison with experiments on turbulence initiated by neutron
absorption in $^3$He-B.\cite{NeutronTurbulence} KH injection which we use
in this work produces almost always more than 3 vortex loops, while in the
neutron absorption the production of 1--2 vortex loops is the most common
case. As has been observed in Ref.~\onlinecite{NeutronTurbulence}, the
transition to turbulence becomes wider towards lower temperatures if it is
initiated by neutron absorption: Even at $0.45\,T_{\rm c}$ regular events
are still rather common.  Numerical simulations of vortex
dynamics\cite{Simulation} also demonstrate that at intermediate
temperatures a single vortex loop in the rotating container may initiate
turbulence or may not, depending on the initial configuration. Also in the flow
of classical liquids in a pipe the transition from laminar to turbulent
flow depends on the initial perturbation: The critical Reynolds number has
been found to scale inversely proportional to the perturbation
amplitude.\cite{PipeFlow}

\begin{figure}[t]
  \centerline{\includegraphics[width=0.7\linewidth]{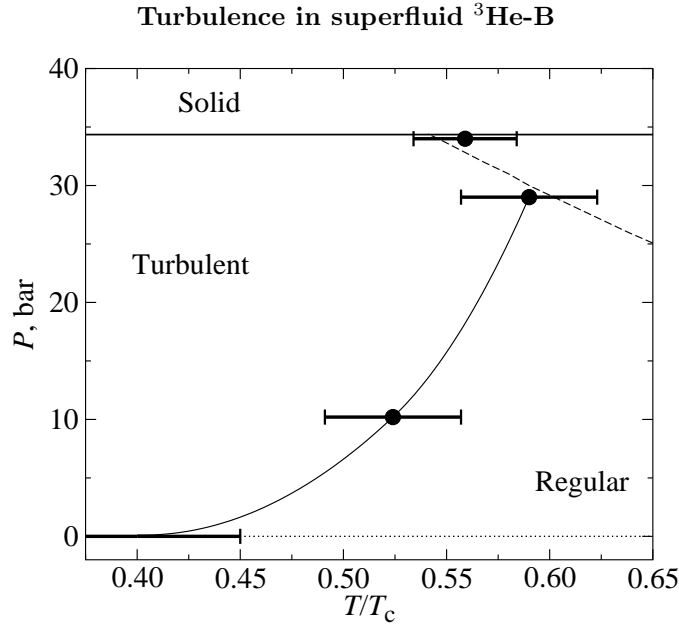}}
\caption{Phase diagram of turbulence in $^3$He-B. The points 
  for 10, 29 and 34 bar represent the middle of the transition $T_{\rm t}$
  from regular dynamics at higher temperatures to turbulent dynamics at
  lower temperatures, as determined from the distributions in
  Fig.~\ref{TurbDistribution}. The error bars indicate the width of the
  transition $\sigma_{\rm t}$ as determined from these distributions (and
  do not indicate the uncertainty in temperature). At zero pressure the
  error bar indicates the unexplored region where the transition to turbulent
  dynamics should occur. The solid line is a guide for the eye. The dashed
  line shows the transition in vortex core structure\cite{CoreTrans} which
  leads to a discontinuity in the mutual friction
  coefficients.\cite{Bevan}
\label{fig:diagr}}
\end{figure}

The dependence of the average transition temperature $T_{\rm t}$ on
pressure is shown in Fig.~\ref{fig:diagr}. The data from
Fig.~\ref{TurbDistribution} have been augmented by the measurements at $P
=0$. At zero pressure we have not observed the turbulence at $T >
0.45\,T_{\rm c}$ at rotation velocities 0.5 -- 0.7\,rad/s. On the other
hand, turbulent vortex tangles have been observed at $T < 0.2\,T_{\rm c}$
using vibrating wires.\cite{AndreevReflection} Combining all the data, we
conclude that there is a pressure dependence of the transition to
turbulence on the $T/T_{\rm c}$ scale.

When inspecting Fig.~\ref{fig:diagr} one should keep in mind that from the
three sets of measurements in Fig.~\ref{TurbTransition-vs-T} more effort
was invested in the temperature calibration and data taking of the runs at
29.0 and 10.2\,bar pressures, while the 34\,bar measurements were more
qualitative in character. Thus, a non-monotonous dependence of $T_{\rm
  t}/T_{\rm c}$ on pressure cannot be claimed. It may simply be an artefact
of the small amount of data measured at 34\,bar. However, we note here an
interesting phenomenon, which might affect the transition to turbulence at
high pressures. As indicated by the dashed line in Fig.~\ref{fig:diagr}
there is a first-order transition\cite{CoreTrans} in the vortex core
structure between a non-axisymmetric core at low $T$, $P$ and an
axisymmetric core at high $T$, $P$. The structure of the core affects the
mutual friction parameters and the dissipation associated with a
non-axisymmetric vortex is larger.\cite{Bevan} It is therefore possible
that there is a kink in the $T_{\rm t}(P)$ dependence at the transition
between the two core structures.

What is the reason for the pressure dependence of the transition to
turbulence? Eq.~\eqref{eq:dyn} suggests that the transition is controlled
by the ratio of the mutual friction parameters, $q = \alpha
/(1-\alpha^{\prime})$. Thus it is instructive to plot our data as a
function of $q$. Unfortunately, the precise measurements of $q(T,P)$ are
available only for two pressures, 10 and 29 bar.\cite{Bevan} Our data for
these pressures from Fig.~\ref{TurbTransition-vs-T} have been repeated as a
function of $q$ in Fig.~\ref{TurbTransition-vs-q}. Generally speaking, the
transition takes place when $q \sim 1$.  However, a critical value of $q$
is not universal: the transition moves to larger $q$ as a function of
pressure. Can we be sure that the difference in $q$ values for the
transitions in Fig.~\ref{TurbTransition-vs-q} is experimentally
significant? First we note that the mutual friction coefficient $\alpha$
can be measured in our experiment in situ using the time-of-flight technique
described in Refs.~\onlinecite{TimeOfFlight} and \onlinecite{LammiProc}.
Such measurements give values of $\alpha$ which agree with
Ref.~\onlinecite{Bevan} both at 29 and 10 bar pressure. Thus there is no
systematic error, for example, in the temperature scales. The scatter of
the data points in Ref.~\onlinecite{Bevan} gives an uncertainty in $q$ of
about $\pm0.04$ at $P=10\,$bar. For $P=29\,$bar the situation is more
difficult to estimate, since there exists an additional scatter due to the
presence of two different types of vortices around $0.6\,T_{\rm c}$. The
possible range of $q$ values at the transition temperature is from
$q\approx1.13$ for the high-temperature axisymmetric vortex to $q \approx
1.72$ for the low-temperature non-axisymmetric vortex (the same type which
occurs at 10 bar). Thus, even with these uncertainties, the critical values
of $q$ for the two pressures in Fig.~\ref{TurbTransition-vs-q} remain well
separated.

Thus, whether the vortex dynamics is turbulent or regular in character, indeed
strongly depends on the magnitude of mutual friction damping. However
the simple classification on the basis of the value of $q$, as suggested in
Ref.~\onlinecite{Turbulence}, agrees with the experiment only as an order of
magnitude estimation. To explain the pressure dependence of the transition to
turbulence presented here more experimental and theoretical work is
required. At the moment we can only speculate about possible reasons.

First, the analysis of Eq.~\eqref{eq:dyn} may not be completely correct.
This equation includes two parameters, $\alpha$ and $\alpha'$, and the
transition may depend on their values in a more complicated way than the
simple ratio $q$. Second, the equation itself may not be fully applicable
to the experimental situation. In particular, in the process of
coarse-graining the mutual friction force over volumes containing many
vortex lines the values of the mutual friction coefficients may become 
renormalized
from their single-vortex values measured in Ref.~\onlinecite{Bevan}. Next,
the properties of the injection of the seed loops may depend on pressure in
such a way that the initial configurations of vortices become less favourable
for initiating the turbulence with decreasing pressure. However, in the
latter case one would expect that the width of the transition will also
increase, similar to the turbulence induced by neutron
absorption,\cite{NeutronTurbulence} which is not observed in
Fig.~\ref{TurbDistribution}.

At present it is assumed that one vital precondition is required to start
turbulence: it is the Kelvin-wave instability\cite{Glaberson} of a vortex
line. It provides the mechanism by which quantized vorticity starts to
multiply even from one single vortex loop which is injected in the rotating
vortex-free counterflow. The Kelvin-wave instability switches on when some
section of the vortex loop becomes oriented along the flow. In this part of
the vortex filament a helical Kelvin wave starts to grow in amplitude. The
wave develops into loops which reconnect to form separated vortex rings.
These rings provide the beginning of the evolving network. The triggering
and development of the Kelvin-wave instability is significantly affected by
the vortex line tension, which has so far been neglected here. The tension
depends, although only logarithmically, on the vortex core size, which
decreases in $^3$He-B by almost an order of magnitude from zero to melting
pressure. This might also influence the transition to turbulence.

The final complication, which we note here, concerns the role of the
magnetic field. Close to the AB interface, where the seed vortex loops are
injected, the magnetic field is high and distorts the superfluid gap in
the B phase as well as the structure of the vortex core. These are the
parameters which directly determine the magnitude of the mutual friction
coefficients.\cite{FrictionTheory} If the turbulence is initiated in the
immediate vicinity of the original position of the seed loops then the
transition to turbulence might also be affected by the pressure and
temperature dependent magnetic field $H_{\rm AB}$. The measurements with
neutrons\cite{NeutronTurbulence} are free from these complications but the
transition in this case is wide, as has been explained above, and the
presently available results do not
allow to pinpoint the critical $q$ value with enough accuracy to judge
its pressure dependence.

\section{CONCLUSIONS}

We have described the first measurements on the phase diagram of turbulence
in $^3$He-B: (i) A wide regime of mutual friction damping has been examined
as a function of temperature, which is not easily accessible in the case of
$^4$He-II. Mutual friction proves to control the character of vortex motion
in superfluid hydrodynamics: when $\alpha/(1-\alpha^{\prime}) \gtrsim 1$
the number of vortex lines is a conserved quantity, while in the opposite
case, $\alpha/(1-\alpha^{\prime}) \lesssim 1$, the vortex number rapidly
proliferates through turbulence. It is known from $^4$He-II that mutual
friction can help to sustain turbulence by blowing up Kelvin-wave
perturbations on vortex loops and it also can damp turbulent motion at
certain length scales.\cite{VinenReview,Kivotides} From $^3$He-B we now
find that high mutual friction can totally suppress turbulence. (ii) The
transition to turbulence is not marked by a unique value of the mutual
friction ratio $q = \alpha/(1-\alpha^{\prime})$; with increasing pressure
the transition moves to higher $q$. (iii) Turbulence is prominently
concentrated in the low-temperature, high-velocity, and high-pressure
corner of the $^3$He-B phase diagram. At higher flow velocities the
transition between regular and turbulent dynamics is velocity independent,
the width of the transition regime is narrow and centered in the
temperature interval 0.5 -- $0.6\,T_{\rm c}$, depending on pressure.

\textbf{Acknowledgments.}
We thank C.~Barenghi, R.~Blaauwgeers, G.~Eska, D.~Kivotides,
N.B.~Kopnin, L.~Skrbek, M.~Tsubota, W.F.~Vinen, and G.E. Volovik 
for valuable discussions. This work has benefited from the EU-IHP ULTI-3
visitor program and the ESF programs COSLAB
and VORTEX.


\vspace{-5mm}
\begin{thebibliography}{99}
\vspace{-1mm}

\bibitem{VinenReview} W.F. Vinen and J.J. Niemela, {\it J. Low Temp. Phys.}
  {\bf 128}, 167 (2002).

\bibitem{Turbulence} A.P. Finne et al., {\it Nature} {\bf 424}, 1022 (2003).

\bibitem{AndreevReflection} S.N. Fisher et al., {\it Phys. Rev. Lett.} {\bf
    86}, 244 (2001).

\bibitem{Bevan} T.D.C. Bevan et al., {\it J. Low Temp. Phys.}, {\bf 109},
  423 (1997). 
  
\bibitem{Sonin} E.B. Sonin, {\it Rev. Mod. Phys.} {\bf 59}, 87 (1987).
  
\bibitem{LammiProc} A.P. Finne et al., {\it J. Low Temp. Phys.}, November
  (2004).

\bibitem{KH} R. Blaauwgeers et al., {\it Phys. Rev. Lett.} {\bf 89}, 155301 (2002).

\bibitem{AB-VortexLayer} R. H\"{a}nninen et al., {\it Phys. Rev. Lett.}
  {\bf 90}, 225301 (2003).

\bibitem{CoreTrans} J.P. Pekola et al., {\it Phys. Rev. Lett.} {\bf 53},
  584 (1984).
  
\bibitem{NeutronTurbulence} A.P. Finne et al., {\it J. Low Temp. Phys.},
  June (2004).
  
\bibitem{Simulation} M. Tsubota and A. Mitani, {\it J. Low Temp. Phys.},
  November (2004).

\bibitem{PipeFlow} B. Hof et al., {\it Phys. Rev. Lett.} {\bf 91}, 244502
  (2003). 
  
\bibitem{TimeOfFlight} A.P. Finne et al., {\it J. Low Temp. Phys.} {\bf
    134}, 375 (2004).
  
\bibitem{Glaberson} R.M. Ostermeier and W.I. Glaberson, {\it J. Low Temp.
    Phys.} {\bf 21}, 191 (1975).

\bibitem{FrictionTheory} N.B. Kopnin, {\it Theory of Nonequilibrium
    Superconductivity} (Oxford University Press, 2001), p. 271.

\bibitem{Kivotides} D.C. Samuels and D. Kivotides, {\it Phys. Rev. Lett.}
  {\bf 83}, 5306 (1999).

\end{thebibliography}
\end{document}